# Multiplicity distribution of colour dipoles at small $x^*$

**G.P. Salam**

*Cavendish Laboratory, Cambridge University,*
*Madingley Road, Cambridge CB3 0HE, UK*

email: salam@hep.phy.cam.ac.uk

**Abstract**

The colour dipole multiplicity distribution is analysed for the wave function of a heavy onium state at small $x$. Numerical results for the average multiplicity and the effect of cutoffs on its power growth are presented. Then, the full multiplicity distribution is analysed: the second multiplicity moment is derived and the tail of the distribution is shown to behave as $\exp(-\log^2 n)$. These results are confirmed by a Monte Carlo simulation which also gives the fluctuations in the spatial density of dipoles.

*Research supported by the UK Particle Physics and Astronomy Research Council

# 1 Introduction

There has been much discussion in the literature relating to the number density of gluons at small $x$ as predicted by the BFKL equation [1, 2, 3]. However until now there has been no analysis of the full multiplicity distribution of gluons. The simplest framework to use is the dipole formalism developed by Mueller [4, 5] to describe the wavefunction of a heavy onium state (for a related approach, see [6]): a heavy onium can emit a gluon at small $x$, which corresponds to one colour dipole ($q\bar{q}$) branching into two ($qg$ and $g\bar{q}$). In the leading $N_C$ approximation each 'child' dipole can then emit gluons independently, resulting in a tree of dipoles. By iterating this process one can determine the wavefunction (squared) for an onium with a given number of gluons, or equivalently, colour dipoles. This framework is formulated in transverse position space. The distribution of dipole position, size and multiplicity plays an important role in high energy onium-onium scattering, particularly for understanding multiple pomeron exchange and the onset of unitarity [7, 8].

In this paper, results for the average multiplicity of gluons are briefly reviewed. Numerical results are presented on the effect of cutoffs on the BFKL power growth and compared with a recent analytical calculation [9]. Three approaches are then taken to determine aspects of the full multiplicity distribution. The first makes use of the dipole generating functional to obtain the second multiplicity moment, $\langle n(n-1)\rangle$, where the generating functional satisfies an integro-differential equation which expresses the branching process. The second approach derives the shape of the high multiplicity tails of the distribution by noting that they result from the production of a single large dipole which cascades into many smaller ones. The third method involves a Monte-Carlo simulation of the branching process, which then allows direct determination of the full multiplicity distribution $P_n$ and also the distribution of fluctuations in the spatial density. The main results are that for intermediate and large sizes, KNO scaling [10, 11, 12] is not observed, but that the tails of the multiplicity distribution are approximately independent of dipole size, showing an $\exp(-\log^2 n)$ behaviour. The fluctuations in density show an exponential decay, whose slope is independent of dipole size and position.

# 2 The generating functional and the first moment

The generating functional $Z(b, y, u)$ is defined such that the probability of finding $n$ dipoles of transverse size $c$ with 'rapidity' ($\log 1/x$) less than $y$, generated from a parent of size $b$ is:

$$P_n(c, b, y) = \frac{1}{n!}\frac{\delta^n Z(b, y, u)}{\delta u(c)^n}|_{u=0}. \tag{1}$$

Defining the $q^{\text{th}}$ multiplicity moment to be $n^{(q)} = \langle n(n-1)\dots(n-q+1)\rangle$, one can see that

$$n^{(q)}(c, b, y) = \frac{\delta^q Z(b, y, u)}{\delta u(c)^q}|_{u=1}. \tag{2}$$

The BFKL equation can then be expressed in the following form [4],

$$\frac{\mathrm{d}Z(b_{01},y,u)}{\mathrm{d}y} = \frac{\alpha_S N_C}{2\pi^2}\int \frac{b_{01}^2 \mathrm{d}^2\mathbf{b}_2}{b_{02}^2 b_{12}^2}\left[Z(b_{02},y,u)Z(b_{12},y,u) - Z(b_{01},y,u)\right], \tag{3}$$

with $\mathbf{b} = \mathbf{b}_{01} = \mathbf{b}_{02} - \mathbf{b}_{12}$. This expresses the branching of the initial dipole, so creating two 'trees' of children and removing the original tree. Functionally differentiating $q$ times with respect to $u(c)$, one obtains an equation for $n^{(q)}$ based on $n^{(0)}$ to $n^{(q-1)}$. With the knowledge that $n^{(0)}$ is 1, this leads to:

$$\frac{\mathrm{d}n^{(q)}(c,b_{01},y)}{\mathrm{d}y} = I^{(q)} + \frac{\alpha_S N_C}{2\pi^2}\int \frac{b_{01}^2 \mathrm{d}^2\mathbf{b}_2}{b_{02}^2 b_{12}^2}[n^{(q)}(c,b_{02},y) + n^{(q)}(c,b_{12},y) - n^{(q)}(c,b_{01},y)], \tag{4}$$

with the inhomogeneous term, $I^{(q)}$, defined to be

$$I^{(q)} = \frac{\alpha_S N_C}{2\pi^2}\int \frac{b_{01}^2 \mathrm{d}^2\mathbf{b}_2}{b_{02}^2 b_{12}^2}\sum_{i=1}^{q-1} C_i^q n^{(i)}(c,b_{02},y) n^{(q-i)}(c,b_{12},y). \tag{5}$$

With these equations one can then solve iteratively for successively higher multiplicity moments. Setting $q=1$, the inhomogeneous term disappears and one obtains and equation for the mean multiplicity whose kernel (in Mellin transform space) is the BFKL kernel [5]. The following solution can be obtained using the saddle point approximation:

$$n^{(1)}(c,b,y) \simeq \frac{b}{c\sqrt{\pi k y}}\exp((\alpha_{\mathcal{P}}-1)y - \log(c/b)^2/ky). \tag{6}$$

This is valid for $|\log(c/b)| \ll ky$. The BFKL power is $(\alpha_{\mathcal{P}}-1) = 4\log 2\alpha_S N_C/\pi$, and $k = 14\alpha_S N_C \zeta(3)/\pi$, with $\zeta(3) \simeq 1.202$ being the Riemann zeta function.

In some of the work that follows, the saddle point solution will not be sufficient, so $n^{(1)}$ has been determined by numerical solution of eq. 4, starting with a single dipole and evolving in rapidity. Results for this are shown in figure 1. In all the calculations presented here, $\alpha_S$ is held fixed at 0.18. As expected, at large rapidity there is good agreement between the saddle point and numerical solution. At lower rapidity (and correspondingly, at very large and small $c$) the numerical solution is significantly different. One reason is that the initial dipole leaves the remainder of a delta-function peak at $c=b$. More significant, is that at large and small $c$, logarithms of size become as important as those of energy and the double leading logarithmic approximation (DLLA) solution should be used:

$$n^{(1)}_{DLLA}(c,b,y) \simeq \left(\frac{\tilde{\alpha}y}{16\pi^2\log^3 b/c}\right)^{1/4}\exp\left[2\sqrt{\tilde{\alpha}y\log\frac{b}{c}}\right], \tag{7}$$

with $\tilde{\alpha} = 2\alpha_S N_C/\pi$. The solution for large sizes is analogous, but with a leading factor of $b^2/c^2$. For $|\log c/b| \gg y$ the DLLA solution deviates only slowly from a power, and is much larger than the saddle-point solution. This is most clearly seen for the $y=4$ curve of figure 1, where the numerical solution gives a relatively straight line, while the saddle point solution drops off very rapidly.

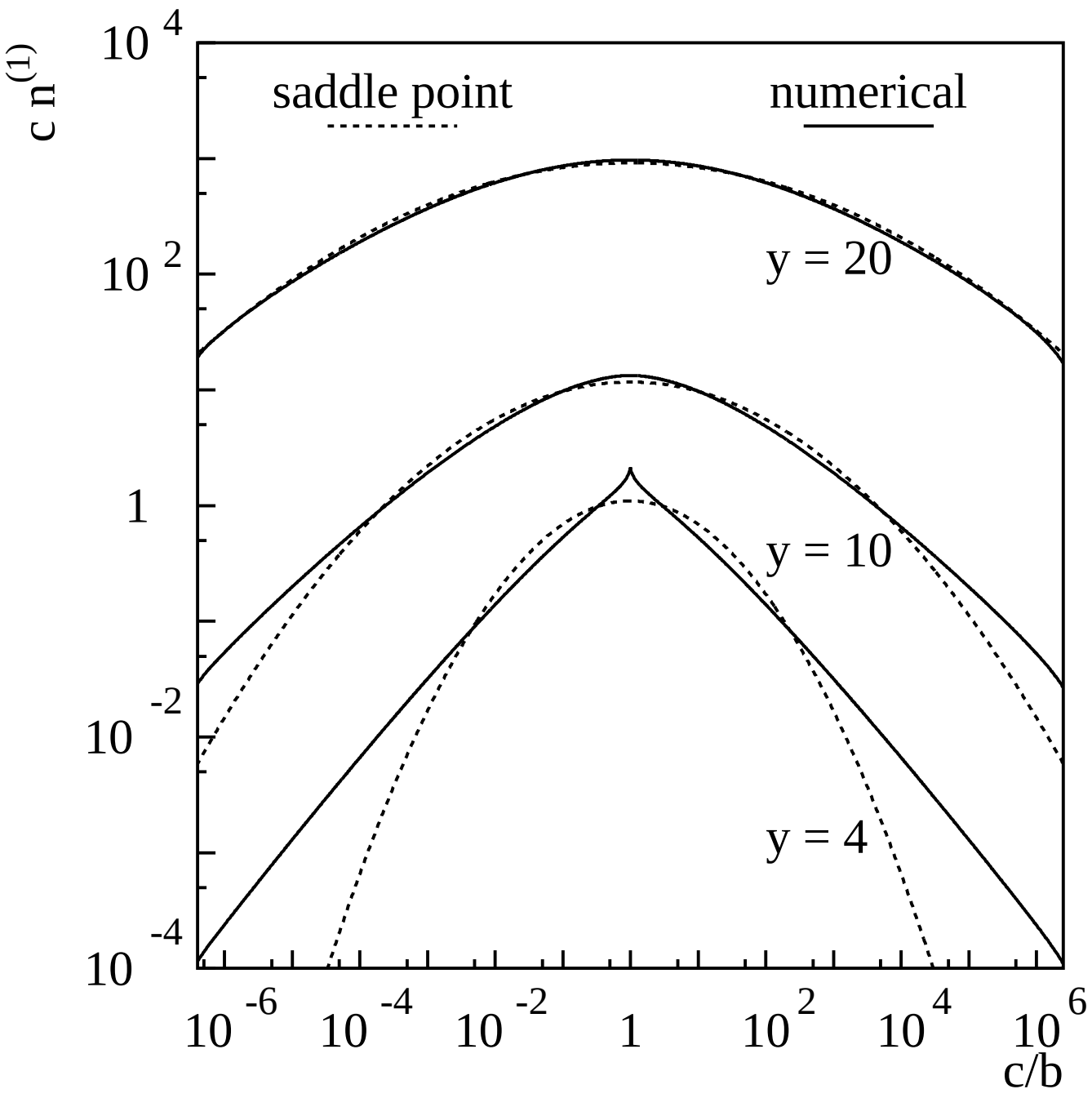

Figure 1: Numerical and saddle point solutions for $cn^{(1)}(c, b, y)$ for three different values of $y$.

Aside from the considering the solution to the full dipole BFKL equation, one can examine the effect of introducing infra-red and ultra-violet cutoffs on transverse size. This relates to recent work which has studied cutoffs in the momentum form of the BFKL equation [13, 14, 15, 9]. An infra-red cutoff is useful to help gauge the effect of eliminating from consideration the non-perturbative region. The justification used for introducing an ultra-violet cutoff is that it gives a crude measure of the effect of energy conservation which places a limit on the largest accessible transverse momentum. Inclusion of running $\alpha_S$ would also provide an effective ultra-violet cutoff [14].

Figure 2 shows the dependence on the ratio of the cutoffs, of the power growth of $n^{(1)}$ with respect to $y$. It also shows the prediction for the leading power from ref. [9].

$$(\alpha_{\mathcal{P}} - 1) \left( \frac{1}{1 + 4\bar{\omega}^2} \right) \tag{8}$$

where $\bar{\omega}$ is the lowest solution to $2\bar{\omega} = \cot[\bar{\omega} \log(c_{max}/c_{min})]$. The shapes of the two solutions appear to be fairly similar, but shifted in $c_{max}/c_{min}$. One could argue that this discrepancy is a consequence of going from cutoffs in position space to ones in momentum space. However any such effect should cancel out when taking the ratio of the cutoffs. Instead, it is probably due to the their scope being different: in ref. [9], they were placed only on the real emission terms of the BFKL kernel, whereas in the numerical solution, they apply to the whole kernel. Note that the virtual terms in the BFKL kernel and in the dipole picture are not equivalent, and that in the dipole picture, it is not possible to put a cutoff on the real terms and not the virtual ones, because it would lead to a divergence.

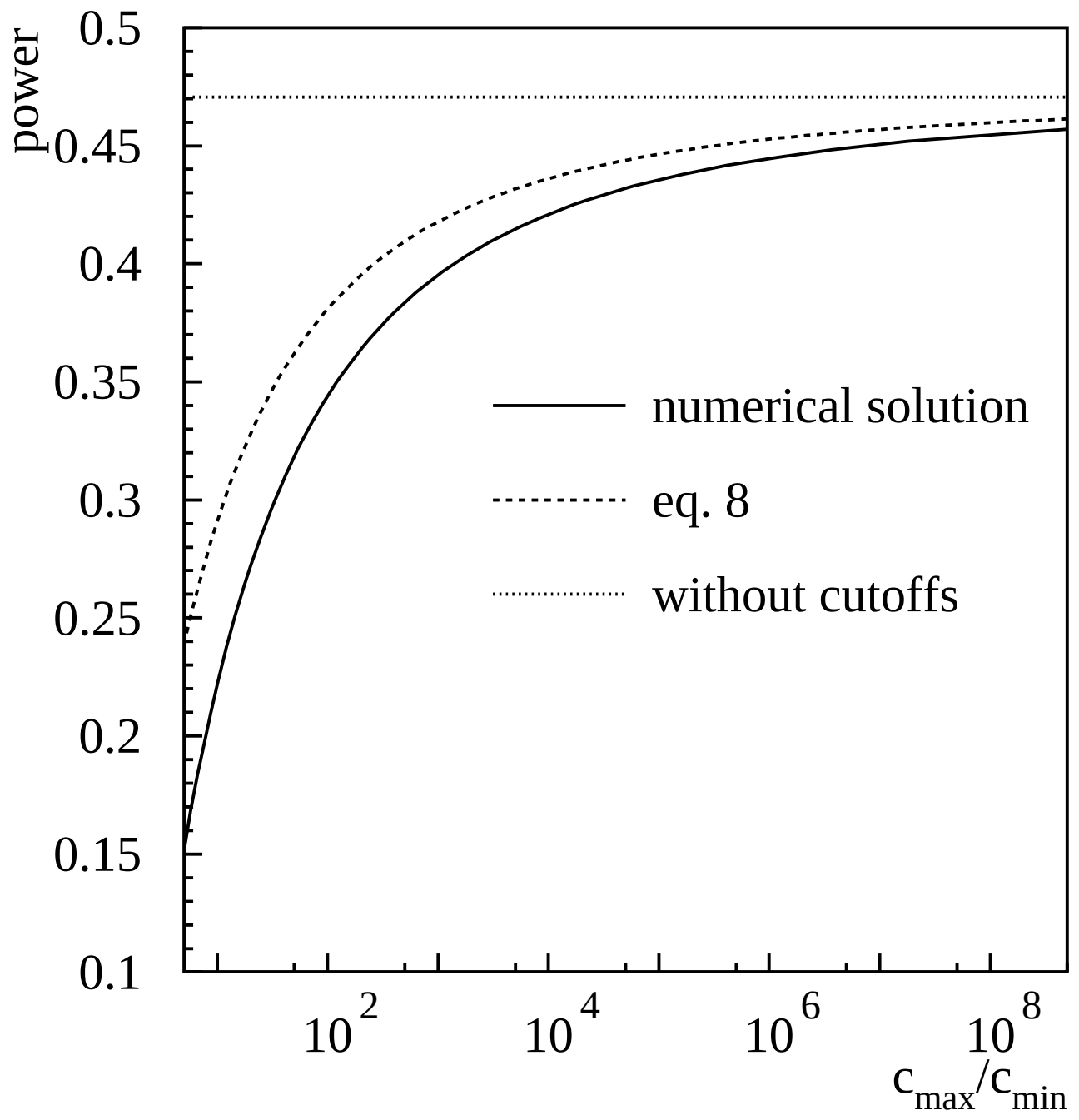

Figure 2: Asymptotic (large $y$) power for the growth of $n^{(1)}$, as a function of the ratio of the upper and lower cutoffs on dipole size. For the curve corresponding to eq. 8, $c_{max}/c_{min}$ should be taken to mean $k_{\perp,max}/k_{\perp,min}$.

When considering the relevance of these calculations to observations at HERA, it should be kept in mind that the power is the asymptotic one. The initial power (ignoring logarithmic corrections) will be the full BFKL power, and the power will decrease only when the width of the solution is of the same order as the separation of the cutoffs.

# 3 The second moment

Once $n^{(1)}$ is known, it can be used to calculate the inhomogeneous term for eq. 4 with $q = 2$. In many other situations where one is solving for higher moments it is possible to use a KNO ansatz, namely that the shape of the probability distribution is independent of the mean,

$$P_n(c,b,y) = \frac{1}{\langle n \rangle} f\left(\frac{n}{\langle n \rangle}\right), \tag{9}$$

and only the mean $\langle n \rangle$ depends on $c$, $b$ and $y$. In terms of multiplicity moments, this is equivalent to saying that:

$$n^{(q)}(c,b,y) \simeq \beta_q [n^{(1)}(c,b,y)]^q, \tag{10}$$

where the $\beta_q$ are independent of $c$, $b$, and $y$. However, when investigating the inhomogeneous term of eq. (4) for the second moment, one finds that the symmetry between large and small scales, which is characteristic of the first moment, is substantially altered, so that a KNO ansatz is not appropriate. This means that the form of the integral for the inhomogeneous term is different for each higher moment.

The evaluation of these integrals is unfortunately quite complicated, and analytical approximations have been found only for a limited ranges of child dipole sizes. It is nonetheless instructive to look at these analytical approximations, because they point to some of the main features of the fluctuations and to the nature of the violations of KNO scaling. Consider $I^{(2)}$, the inhomogeneous term for the second moment.

$$I^{(2)} \sim \int \frac{b_{01}^2 \mathrm{d}^2 \mathbf{b}_2}{b_{02}^2 b_{12}^2} n^{(1)}(c, b_{02}, y) n^{(1)}(c, b_{12}, y). \tag{11}$$

For large values of $c/b_{01}$ one can make the approximation that $b_{02} \simeq b_{12}$ in the dominant region of the integration. The two exponents in the integrand are then the same and easily integrated over, giving

$$I^{(2)} \sim \frac{\mathrm{e}^{2(\alpha_{\mathcal{P}}-1)y}}{\sqrt{2\pi k y}} \left(\frac{b_{01}}{c}\right)^2, \qquad (c \gg b_{01}). \tag{12}$$

This is valid for $\log(c/b_{01}) \gg \sqrt{ky}$, though it remains a reasonable approximation down to values of $c$ close to $b_{01}$ and remains true even outside the saddle point approximation for $n^{(1)}$. The only other region where the inhomogeneous term can be estimated is for small $c$. In this case, most of the contribution to the integral comes from the regions of small $b_{02}$ and small $b_{12}$, where one of the exponentials decouples, leaving only the other to integrate over. The main difference is that the leading powers of $c/b$ shift the position of the maximum of the integrand, and in the process introduce an extra exponential dependence on the rapidity. This leaves $I^{(2)}$ with the following dependence for small $c$:

$$I^{(2)} \simeq \frac{b_{01} \mathrm{e}^{2(\alpha_{\mathcal{P}}-1)y}}{c\sqrt{\pi k y}} \mathrm{e}^{ky/4 - \log(b_{01}/c)^2/ky}, \qquad (c \ll b_{01}). \tag{13}$$

More precisely, this is valid for $-ky \ll \log(c/b_{01}) \ll -ky/2 - \sqrt{ky}$. Violation of KNO as transverse size $c$ is varied, arises because of the extra exponential dependence on $y$, which is present for small $c$ but not large $c$. As a result $I^{(2)}$ has its maximum at a smaller value of transverse size than $n^{(1)}$. To a rough approximation, the maximum is shifted from $c = b$ for $cn^{(1)}$ to $c \simeq ky/4$ for $cI^{(2)}$.

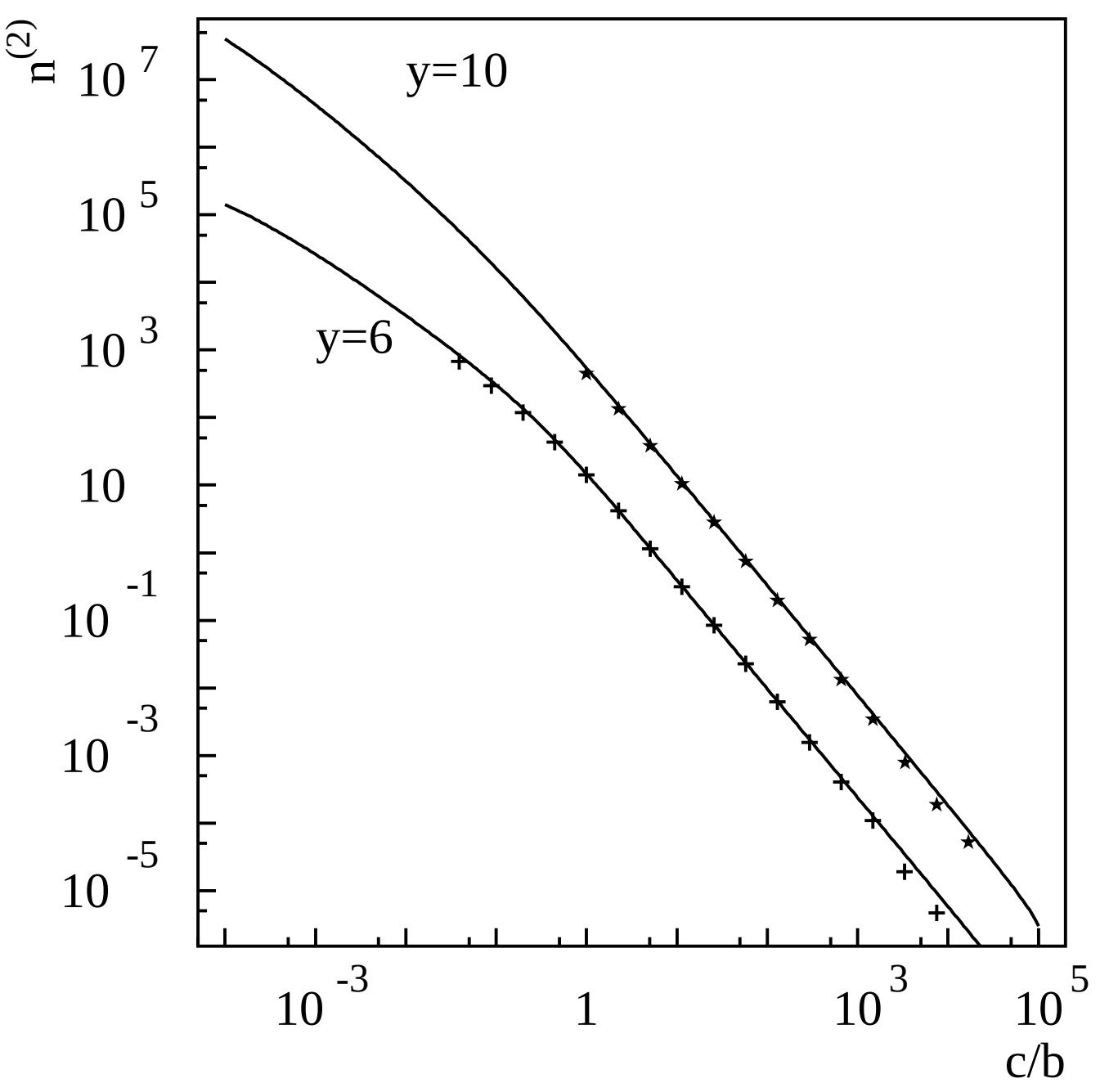

Figure 3: The second multiplicity moment as a function of dipole size. The curves are generated by numerical solution of eq. (4) for $n^{(2)}$. The points correspond to Monte Carlo results where the bin includes sizes up to a factor of 1.5 either side of the central value. For $y = 6$ a lower cutoff of $0.01b_0$ is used, while for $y = 10$ the lower cutoff is $0.1b_0$.

Having established the form of $I^{(2)}$, one can examine $n^{(2)}$ which is related to $I^{(2)}$ by eq. 4. Figure 3 was obtained by numerical solution of eq. (4) for $q = 2$. There are various stages to the numerical determination of $n^{(2)}$: the first is to take the full numerical solution for $n^{(1)}$ and fit it with a simple analytical approximation having the correct asymptotic behaviour. The fit is then used to evaluate the integral for the inhomogeneous term as a function of $c$. Finally, this is fed in as a trial solution to eq. (4), the actual solution being obtained iteratively.

The main features of $I^{(2)}$ can be seen to be also present in $n^{(2)}$: the $c^{-2}$ behaviour at large $c$ is reflected in the steep descent in $n^{(2)}$ (though the actual power in the region shown is slightly larger than $-2$). At smaller dipole sizes there is a tailing off of $n^{(2)}$, as expected, and this tailing off occurs at larger $c$ for $y = 6$ than for $y = 10$.

When trying to understand the shape (scaled by the mean) of the probability distri-bution $P_n(c,b,y)$ it is the normalised moments $\beta_q = n^{(q)}/[n^{(1)}]^q$ which are of relevance[2]. If one considers $\beta_2$ for large $c$, assuming that $n^{(2)}(c) \propto I^{(2)}(c)$, one finds that:

$$\beta_2 \sim \mathrm{e}^{2\log(c/b)^2/ky}. \tag{14}$$

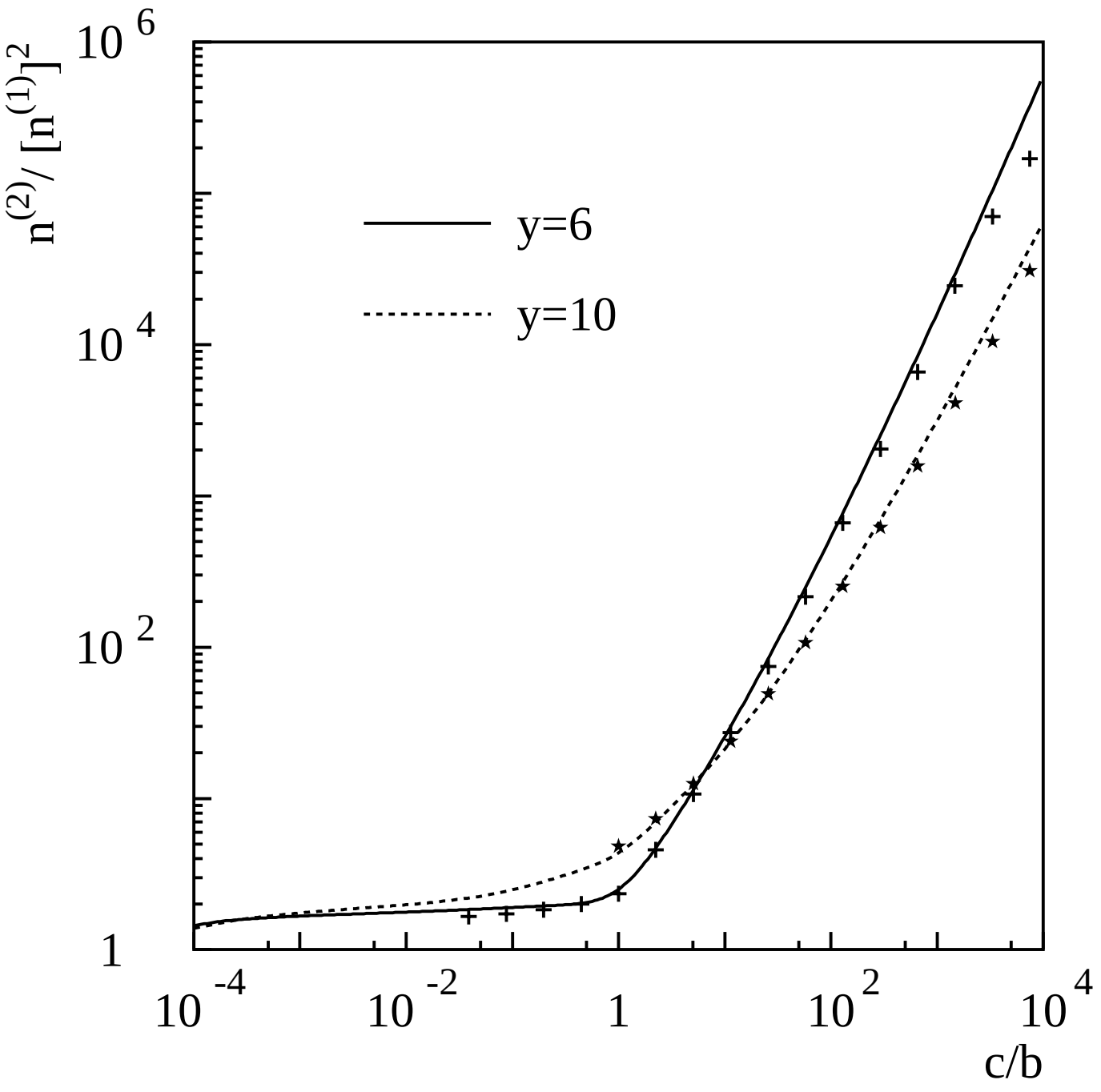

Figure 4: The curves correspond to $n^{(2)}/[n^{(1)}]^2$ as obtained by solving eq. (4) numerically. The points are results from Monte Carlo simulations of the branching process.

This corresponds to the probability distribution $P_n(c,b,y)$ being much broader relative to its mean for large sizes of dipoles than for intermediate sizes. This is to be expected: though the probability of producing a single large dipole is relatively small, once this has occurred, many more are quite likely to follow, increasing the values of all the higher moments.

Figure 4 shows the dependence of $\beta_2 = n^{(2)}/[n^{(1)}]^2$ on dipole size. It is clear that at large sizes there is a very rapid increase in $\beta_2$. As $y$ increases, the steepness of the curve for $\beta_2$ is reduced, in accordance with expectations from eq. (14). At very small sizes, $\beta_2$ is almost independent of both rapidity and size. This is a signature of KNO scaling and certainly not consistent with one's expectations from eq. 13. The reason for this is that the validity of eq. 13 is limited by the range of applicability of the BFKL saddle point solution. Instead one should use the DLLA solution. This is known to give KNO scaling with the following relations for the coefficients $\beta_q$ [11]:

$$\beta_q = \frac{q^2}{2(q^2-1)} \sum_{i=1}^{q-i} C_i^q \frac{\beta_i \beta_{q-i}}{i(q-i)}. \tag{15}$$

[2]Strictly speaking, one should use $\langle n^q \rangle$ and opposed to $n^{(q)} = \langle n(n-1)\ldots(n-q+1)\rangle$, however this will only have an effect where $n^{(q)}$ is small, i.e. for large values of $\log(c/b)$. Then, using $n^{(q)}$ instead $\langle n^q\rangle/n^{(1)}$ will cause one to underestimate the width of the distribution.

This is only valid in the region of asymptotically small size. One can see that the curves in figure 4 are approaching the value of $\beta_2 = 4/3$ from eq. 15.

So far we have considered in detail only the first and second multiplicity moments. It would be of interest to obtain more information about higher moments, to help understand the tail of the distribution. In principle, eq. (4) could be solved numerically for successively higher moments, using the previously determined lower moments to construct the inhomogeneous terms. This has in fact been done for the third moment, but is not feasible for arbitrarily high moments. An alternative approach is presented in the next section.

# 4 The tail of the multiplicity distribution

The tail of the distribution can be analysed with the assumption that the occasional high multiplicity event results from the initial production of a large 'source' dipole which produces a cascade of smaller dipoles. The probability of producing a large dipole of size $a$ is proportional to $1/a^2$. On average it will then produce $n^{(1)}_{DLLA}(c,a,y)$ dipoles. From eq. 7 therefore, the size $a$ associated with a multiplicity $n$ is $a \sim c\exp[\log^2 n/4\tilde{\alpha}y]$, giving a probability $P_n \sim \exp[-\log^2 n/2\tilde{\alpha}y]/c^2$. This indicates a distribution with a very long tail which is independent of dipole size except that its normalisation scales as $1/c^2$.

The distribution can be calculated in more detail using the moments for the DLLA distribution produced by the large dipole:

$$n^{(q)}_{DLLA} = \beta_q [n^{(1)}_{DLLA}]^q, \tag{16}$$

with (from eq. 15)

$$\beta_q \simeq 2\gamma^q q q!, \tag{17}$$

where $\gamma$ is 0.39 [11]. To obtain the moment of the actual multiplicity distribution, the DLLA moment for a given source size $a$ is multiplied by the probability of producing a dipole of that size, and the result is integrated over $a$:

$$n^{(q)}(c,y) \sim \int_c^\infty \frac{\mathrm{d}\log a}{a^2} n^{(q)}_{DLLA}(c,a,y). \tag{18}$$

It would be more accurate to include the DLLA correction to the probability for producing a large dipole. But this is a correction of order $O(1/q)$, so it will have little effect on the shape of the tail, though it will affect the normalisation, which is not being determined here. To obtain $n^{(q)}$, eq. 18 is evaluated with the saddle point approximation. The logarithmic prefactor in eq. 7 is held constant when determining the position of the saddle point, which is again equivalent to taking the limit of large $q$. The result is then integrated with respect to the rapidity of the initially produced dipole. This gives

$$n^{(q)} \sim \frac{1}{c^2}\left(\frac{\phi}{q}\right)^{q/2} q^{1/2} e^{q^2\tilde{\alpha}y/2}, \tag{19}$$

with $\phi$ being

$$\phi = \frac{2\gamma^2 w^2}{\pi\tilde{\alpha}y e^2}. \tag{20}$$

The width of the bin for $c$ is $w = \log(c_{max}/c_{min})$. The normalisation of the moments has been left out because it would in any case be altered by inclusion of the DLLA corrections to the probability of producing the initial dipole. The probability distribution for large $n$ is obtained from the moments, by taking the inverse Mellin transform of $n^{(q)}$, using the saddle point approximation to give

$$P_n(c,y) \sim \frac{1}{c^2}\exp\left[-\frac{\log^2 n}{2\tilde{\alpha}y} - \log n\left(1 + \frac{1}{2\tilde{\alpha}y}\log\left(\frac{\log n}{\tilde{\alpha}y\phi}\right)\right) + \frac{1}{2}\log\left(\frac{\log n}{\tilde{\alpha}y}\right)\right] \tag{21}$$

This is similar to the expression derived earlier, but also includes corrections relevant to moderate multiplicities.

# 5 Monte Carlo calculation of the multiplicity distribution

One way of testing these ideas is to use a Monte Carlo program to simulate the dipole branching process: this gives access to all exclusive information on the dipole distribution in rapidity and transverse position, from which one can directly determine the multiplicity distribution as well as the moments.

There are various limitations inherent to the Monte Carlo approach. First is the issue of the lower cutoff, which is needed to limit the number of dipoles being produced: because of a divergence in the kernel of the BFKL equation, the total number of dipoles being generated for a given logarithmic range of dipole sizes varies roughly as $n^{(1)} \simeq b/c$. This makes it impossible to study the multiplicity moments at small values of $c$, as well as distorting the behaviour of the moments close to the cutoff. The results for physical quantities such as cross sections should, essentially, be independent of this single cutoff [4, 8]. One is also limited in the maximum rapidity that can be studied, since at large $y$, large numbers of dipoles are produced per event, again increasing the time needed to generate an event. Finally, to be able to examine the tails of the distributions one needs high statistics, especially at large dipole sizes where probabilities are suppressed by a factor of $c^{-2}$.

In figures 3 and 4, results from the Monte Carlo program are shown together with those from numerical solution of eq. (4). There is very good agreement between the two approaches, though at larger dipole sizes the Monte Carlo results are slightly low: this is a consequence of insufficient statistics in the region of the multiplicity distribution contributing most to the second moment.

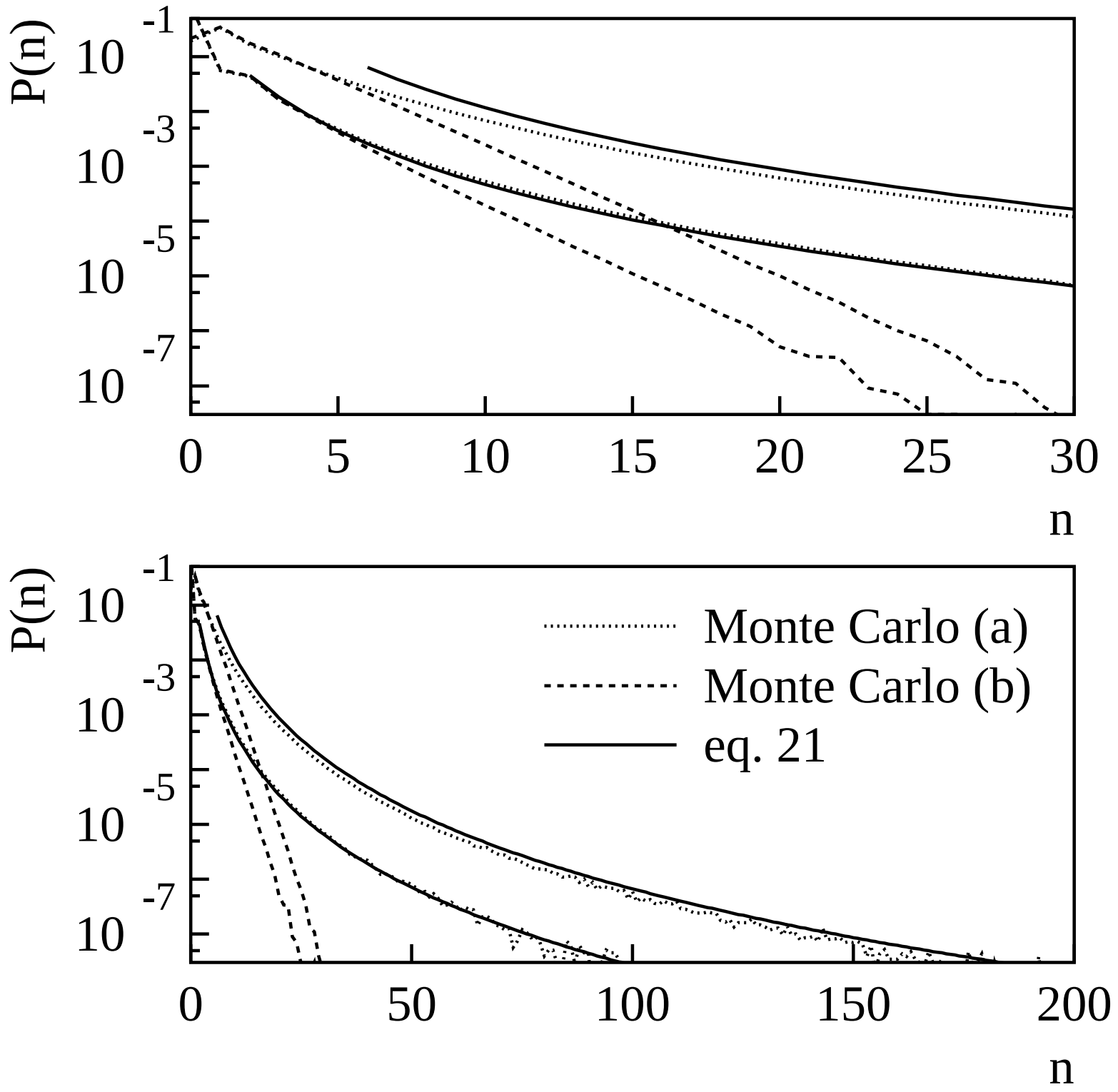

Figure 5: The Monte Carlo curve (a) is the probability distribution for the total multiplicity of dipoles of size $c \simeq c_0$. The curves (b) are for the multiplicity in a central disc of radius $2c_0$. In each plot, the upper set of curves corresponds to $c_0 = b$, the lower to $c_0 = 5b$. (Apart from the scale for $n$, the upper and lower plots are identical)

Figure 5 shows the multiplicity distribution as determined directly with the Monte Carlo simulation (curve (a)), together with the asymptotic prediction from eq. 21, for two ranges of dipole size. When plotting the theoretically predicted distribution it is necessary to fix the normalisation. This is done so as to give agreement with the Monte Carlo results for larger sizes. The shape of the predicted distribution agrees surprisingly well with that of the Monte Carlo results, even down to a multiplicity as low as 2. For the smaller sizes, the prediction is high, particularly at lower multiplicities. The reason for this, is that the probability of producing the initial 'large' dipole of size $a$ falls below the $1/a^2$ approximation, especially at the smaller values of $a$ which are relevant to low multiplicities.

The curves (b) from the Monte Carlo simulation are very different. They shows the distribution of the number in a restricted region of transverse position space: for a dipole to contribute, its centre must be less than $2c_0$ from the centre of the source dipole (where $c_0$ is roughly the mean of the dipole sizes being considered). This gives a measure of the fluctuations in the spatial density, which may be more relevant to multiple pomeron exchange than the fluctuations in the total multiplicity [7, 8]. The distribution of densities shows an exponential decrease, very distinct from the $\exp(-\log^2 c)$ behaviour of the total multiplicity. This difference arises because the density of dipoles from a large initial source is controlled by two factors: the total number of dipoles $\sim \exp(\sqrt{\bar{\alpha} y} \log a)$, divided by the total area which scales as $a^{-2} = \exp(-2 \log a)$, which means that the occasional large source will not contribute to high densities.

For a region of size $r$, the largest numbers of dipoles within the region will therefore be produced by an initial source no larger than $r$. Because $r$ is not much bigger than $c_0$ it is not really legitimate to apply the DLLA approximation to determine the fluctuations. If one nevertheless tries, one obtains a multiplicity distribution proportional to $n \exp(-n/\mu_{DLLA})$, with

$$\mu_{DLLA} \simeq \gamma n^{(1)}_{DLLA}(r, c, y). \tag{22}$$

Fitting a similar distribution to the Monte Carlo curves (b) one can obtain a value $\mu_{MC}$ and compare the two, as in the following table:

| $y$ | $\mu_{DLLA}$ | $\mu_{MC}$ |
|---|---|---|
| 6 | 1.52 | 1.65 |
| 10 | 3.45 | 4.3 |

For $y = 6$ there is a reasonable agreement, but for $y = 10$ there is already a significant deviation, which arises because the DLLA solution is not appropriate for such a large rapidity with such a small range of size. It is therefore necessary to include the full BFKL fluctuations. Work on this is in progress.

The only other property which has been derived for the fluctuations in density, is that for a given rapidity, and ignoring normalisation, the tails of the distribution should be independent of the size of dipole and the transverse position (as long as the size of the region being studied is proportional to the dipole size). This can be shown as follows: there is a configuration which, for a particular size and location, favours the largest fluctuations after evolution by rapidity $y$. By a suitable sequence of branching, this configuration can be produced at low rapidity, anywhere in transverse position and at any scale, after which it has the whole rapidity range $y$ available to generate the large fluctuations at the new size and location. Therefore the tail of the density distribution will be independent of position and dipole size.

This is seen in the lower plot of figure 5, where the distributions for the two sizes of dipole (curves (b)) have almost identical slopes. When examining the distribution at different positions, one is quickly limited by statistics. However, at sufficiently large multiplicities the tails do also seem to be exponential, again with the same slope.

# 6 Conclusions

In this paper a variety of aspects of the heavy onium small-$x$ dipole distribution have been studied. Aspects of the average multiplicity were reviewed and numerical results on the effect of transverse size cutoffs on the asymptotic power growth were presented. Three approaches were then taken to analyse the fluctuations in the multiplicity. The first made use of the generating functional to obtain equations for the multiplicity moments. Some of the features of these moments were obtained analytically and the full numerical solution for $n^{(2)}$ was also presented. The second approach analysed the tails of the multiplicity distributions with the assumption that high multiplicities are caused by a cascade down from a single large dipole, giving an $\exp(-\log^2 n/2\tilde{\alpha}y)$ behaviour. Thirdly, the branching process was simulated with a Monte Carlo program, confirming both the details of the second multiplicity moment and the shape of the tails of the multiplicity distribution, but showing that fluctuations in spatial density have an exponential distribution.

Apart from their intrinsic interest, the main importance of these results is in understanding multiple pomeron exchange, which is associated with the onset of unitarity corrections in onium-onium scattering [7]. Consider two pomeron exchange. This is sensitive to high-multiplicity configurations. Because of the long tail of the multiplicity distribution, it should be enhanced. But at moderate rapidities, this enhancement comes only from a small fraction of the configurations (the tail), while single pomeron exchange will be dominated by the majority of configurations, those of low multiplicity. Therefore two pomeron exchange might appear to be significant, but it will actually give unitarity corrections in only a small fraction of events. To gain a better understanding of unitarity corrections, one should take into account all numbers of pomeron exchanges, as suggested in [7]. A Monte Carlo analysis based on this approach will be presented elsewhere [8].

It might also be possible to observe large fluctuations in multiplicities at HERA, for example in the rapidity region between the hard process and the proton remnant. This would be an indirect consequence of the enhancement of multiple pomeron exchange, because cutting through $n$ pomerons should increase the multiplicity by a factor $n$.

# Acknowledgements

I am very grateful to B.R. Webber and A.H. Mueller for suggesting this work and for useful comments.

## References

[1] Y. Y. Balitskiĭ and L. N. Lipatov, Sov. Phys. JETP **28**, 822 (1978).

[2] E. A. Kuraev, L. N. Lipatov, and V. S. Fadin, Sov. Phys. JETP **45**, 199 (1977).

[3] L. N. Lipatov, Sov. Phys. JETP **63**, 904 (1986).

[4] A. H. Mueller, Nucl. Phys. **B415**, 373 (1994).

[5] A. H. Mueller and B. Patel, Nucl. Phys. **B425**, 471 (1994).

[6] N. N. Nikolaev, B. G. Zakharov, and V. R. Zoller, JETP Lett. **59**, 6 (1994).

[7] A. H. Mueller, Nucl. Phys. **B437**, 107 (1995).

[8] G. P. Salam, Cavendish-HEP preprint, in preparation .

[9] M. F. McDermott, J. R. Forshaw, and G. G. Ross, CERN-TH. 7552/95 (1995).

[10] Z. Koba, H. B. Nielsen, and P. Olesen, Nucl. Phys. **40**, 317 (1972).

[11] Y. L. Dokshitzer, V. A. Khoze, A. H. Mueller, and S. I. Troyan, *Basics of Perturbative QCD*, chapter 5, pages 132–138, Editions Frontières, 1991.

[12] Y. L. Dokshitzer, Phys. Lett. **B 305**, 295 (1993).

[13] J. C. Collins and P. V. Landshoff, Phys. Lett. **B 276**, 196 (1992).

[14] A. Askew, J. Kwieciński, A. Martin, and P. Sutton, Phys. Rev. **D49**, 4402 (1994).

[15] J. R. Forshaw, P. N. Harriman, and P. J. Sutton, Nucl. Phys. **B416**, 739 (1994).